\documentclass[conference]{IEEEtran}
\IEEEoverridecommandlockouts
\usepackage{cite}
\usepackage{amsmath,amssymb,amsfonts}
\usepackage{graphicx}

\usepackage{textcomp}
\usepackage{xcolor}
\def\BibTeX{{\rm B\kern-.05em{\sc i\kern-.025em b}\kern-.08em
    T\kern-.1667em\lower.7ex\hbox{E}\kern-.125emX}}

\usepackage{amsmath,amsfonts}

\usepackage{array}

\usepackage[font=small]{caption}
\usepackage{subcaption}
\usepackage{textcomp}
\usepackage{stfloats}
\usepackage{url}
\usepackage{verbatim}
\usepackage{graphicx}

\usepackage{url}
\usepackage[utf8]{inputenc}
\usepackage{xcolor}
\usepackage{xspace}
\usepackage{epsfig}
\usepackage{balance}
\usepackage{booktabs}
\usepackage{tabularx}
\usepackage{cite}
\usepackage{mathtools}
\usepackage{soul}
\usepackage{lipsum}
\usepackage{bm}
\usepackage{enumerate}
\usepackage{enumitem}

\usepackage[sort&compress,numbers]{natbib}

\usepackage[acronyms,nonumberlist,nopostdot,nomain,nogroupskip,acronymlists={hidden}]{glossaries}
\newglossary[algh]{hidden}{acrh}{acnh}{Hidden Acronyms}
\glsdisablehyper
\newacronym{3gpp}{3GPP}{3rd Generation Partnership Project}
\newacronym{4g}{4G}{4th generation mobile network}
\newacronym{5g}{5G}{5th generation mobile network}
\newacronym{6g}{6G}{6th generation mobile network}
\newacronym{nextg}{NextG}{Next Generation}
\newacronym{5gc}{5GC}{5G Core}
\newacronym{adc}{ADC}{Analog to Digital Converter}
\newacronym{aerpaw}{AERPAW}{Aerial Experimentation and Research Platform for Advanced Wireless}
\newacronym{ai}{AI}{Artificial Intelligence}
\newacronym{aimd}{AIMD}{Additive Increase Multiplicative Decrease}
\newacronym{am}{AM}{Acknowledged Mode}
\newacronym{amc}{AMC}{Adaptive Modulation and Coding}
\newacronym{amf}{AMF}{Access and Mobility Management Function}
\newacronym{aops}{AOPS}{Adaptive Order Prediction Scheduling}
\newacronym{api}{API}{Application Programming Interface}
\newacronym{apn}{APN}{Access Point Name}
\newacronym{aqm}{AQM}{Active Queue Management}
\newacronym{ausf}{AUSF}{Authentication Server Function}
\newacronym{avc}{AVC}{Advanced Video Coding}
\newacronym{awgn}{AGWN}{Additive White Gaussian Noise}
\newacronym{balia}{BALIA}{Balanced Link Adaptation Algorithm}
\newacronym{bbu}{BBU}{Base Band Unit}
\newacronym{bdp}{BDP}{Bandwidth-Delay Product}
\newacronym{ber}{BER}{Bit Error Rate}
\newacronym{bf}{BF}{Beamforming}
\newacronym{bler}{BLER}{Block Error Rate}
\newacronym{brr}{BRR}{Bayesian Ridge Regressor}
\newacronym{bsr}{BSR}{Buffer Status Report}
\newacronym{bs}{BS}{Base Station}
\newacronym{bpsk}{BPSK}{Binary Phase-shift keying}
\newacronym{bss}{BSS}{Business Support System}
\newacronym{ca}{CA}{Carrier Aggregation}
\newacronym{caas}{CaaS}{Connectivity-as-a-Service}
\newacronym{cb}{CB}{Code Block}
\newacronym{cc}{CC}{Congestion Control}
\newacronym{ccid}{CCID}{Congestion Control ID}
\newacronym{cco}{CC}{Carrier Component}
\newacronym{cd}{CD}{Continuous Delivery}
\newacronym{cdd}{CDD}{Cyclic Delay Diversity}
\newacronym{cdf}{CDF}{Cumulative Distribution Function}
\newacronym{cdma}{CDMA}{Code-Division Multiple Access}
\newacronym{cdn}{CDN}{Content Distribution Network}
\newacronym{ci}{CI}{Continuous Integration}
\newacronym{cicd}{CI/CD}{Continuous Integration/Continuous Delivery}
\newacronym{cir}{CIR}{Channel Impulse Response}
\newacronym{cn}{CN}{Core Network}
\newacronym{codel}{CoDel}{Controlled Delay Management}
\newacronym{comac}{COMAC}{Converged Multi-Access and Core}
\newacronym{cord}{CORD}{Central Office Re-architected as a Datacenter}
\newacronym{cornet}{CORNET}{COgnitive Radio NETwork}
\newacronym{cosmos}{COSMOS}{Cloud Enhanced Open Software Defined Mobile Wireless Testbed for City-Scale Deployment}
\newacronym{cots}{COTS}{Commercial Off-the-Shelf}
\newacronym{cp}{CP}{Control Plane}
\newacronym{cpu}{CPU}{Central Processing Unit}
\newacronym{cqi}{CQI}{Channel Quality Information}
\newacronym{cr}{CR}{Cognitive Radio}
\newacronym{cran}{CRAN}{Cloud \gls{ran}}
\newacronym{crs}{CRS}{Cell Reference Signal}
\newacronym{csi}{CSI}{Channel State Information}
\newacronym{csirs}{CSI-RS}{Channel State Information - Reference Signal}
\newacronym{cu}{CU}{Central Unit}
\newacronym{d2tcp}{D$^2$TCP}{Deadline-aware Data center TCP}
\newacronym{d3}{D$^3$}{Deadline-Driven Delivery}
\newacronym{dac}{DAC}{Digital to Analog Converter}
\newacronym{dag}{DAG}{Directed Acyclic Graph}
\newacronym{darpa}{DARPA}{Defense Advanced Research Projects Agency}
\newacronym{das}{DAS}{Distributed Antenna System}
\newacronym{dash}{DASH}{Dynamic Adaptive Streaming over HTTP}
\newacronym{dbs}{DBS}{Deep Brain Stimulation}
\newacronym{dc}{DC}{Dual Connectivity}
\newacronym{dccp}{DCCP}{Datagram Congestion Control Protocol}
\newacronym{dce}{DCE}{Direct Code Execution}
\newacronym{dci}{DCI}{Downlink Control Information}
\newacronym{dcl}{DCL}{Dear Colleague Letter}
\newacronym{dctcp}{DCTCP}{Data Center TCP}
\newacronym{devops}{DevOps}{Development and Operations}
\newacronym{dl}{DL}{Deep Learning}
\newacronym{dmr}{DMR}{Deadline Miss Ratio}
\newacronym{dmrs}{DMRS}{DeModulation Reference Signal}
\newacronym{drlcc}{DRL-CC}{Deep Reinforcement Learning Congestion Control}
\newacronym{drs}{DRS}{Discovery Reference Signal}
\newacronym{dt}{DT}{Digital Twin}
\newacronym{dtn}{DTN}{Digital Twin Network}
\newacronym{dtmn}{DTMN}{Digital Twin for Mobile Network}
\newacronym{dtwn}{DTWN}{Digital Twin Wireless Network}
\newacronym{du}{DU}{Distributed Unit}
\newacronym{e2e}{E2E}{end-to-end}
\newacronym{ecaas}{ECaaS}{Edge-Cloud-as-a-Service}
\newacronym{ecn}{ECN}{Explicit Congestion Notification}
\newacronym{edf}{EDF}{Earliest Deadline First}
\newacronym{em}{EM}{Electro-Magnetic}
\newacronym{embb}{eMBB}{Enhanced Mobile Broadband}
\newacronym{empower}{EMPOWER}{EMpowering transatlantic PlatfOrms for advanced WirEless Research}
\newacronym{enb}{eNB}{evolved Node Base}
\newacronym{endc}{EN-DC}{E-UTRAN-\gls{nr} \gls{dc}}
\newacronym{epc}{EPC}{Evolved Packet Core}
\newacronym{eps}{EPS}{Evolved Packet System}
\newacronym{es}{ES}{Edge Server}
\newacronym{etsi}{ETSI}{European Telecommunications Standards Institute}
\newacronym[firstplural=Estimated Times of Arrival (ETAs)]{eta}{ETA}{Estimated Time of Arrival}
\newacronym{eutran}{E-UTRAN}{Evolved Universal Terrestrial Access Network}
\newacronym{faas}{FaaS}{Function-as-a-Service}
\newacronym{fapi}{FAPI}{Functional Application Platform Interface}
\newacronym{fcc}{FCC}{Federal Communications Commission}
\newacronym{fdd}{FDD}{Frequency Division Duplexing}
\newacronym{fdm}{FDM}{Frequency Division Multiplexing}
\newacronym{fdma}{FDMA}{Frequency Division Multiple Access}
\newacronym{fed4fire}{FED4FIRE+}{Federation 4 Future Internet Research and Experimentation Plus}
\newacronym{fir}{FIR}{Finite Impulse Response}
\newacronym{fit}{FIT}{Future \acrlong{iot}}
\newacronym{fl}{FL}{Federated Learning}
\newacronym{fpga}{FPGA}{Field Programmable Gate Array}
\newacronym{fr2}{FR2}{Frequency Range 2}
\newacronym{fs}{FS}{Fast Switching}
\newacronym{fscc}{FSCC}{Flow Sharing Congestion Control}
\newacronym{ftp}{FTP}{File Transfer Protocol}
\newacronym{fw}{FW}{Flow Window}
\newacronym{ga128}{Ga}{Golay Sequence type A}
\newacronym{ge}{GE}{Gaussian Elimination}
\newacronym{glfsr}{GLFSR}{Galois Linear Feedback Shift Register}
\newacronym{gnb}{gNB}{Next Generation Node Base}
\newacronym{gold}{Gold}{Gold}
\newacronym{gop}{GOP}{Group of Pictures}
\newacronym{gpr}{GPR}{Gaussian Process Regressor}
\newacronym{gpu}{GPU}{Graphics Processing Unit}
\newacronym{gtp}{GTP}{GPRS Tunneling Protocol}
\newacronym{gtpc}{GTP-C}{GPRS Tunnelling Protocol Control Plane}
\newacronym{gtpu}{GTP-U}{GPRS Tunnelling Protocol User Plane}
\newacronym{gtpv2c}{GTPv2-C}{\gls{gtp} v2 - Control}
\newacronym{gw}{GW}{Gateway}
\newacronym{harq}{HARQ}{Hybrid Automatic Repeat reQuest}
\newacronym{hetnet}{HetNet}{Heterogeneous Network}
\newacronym{hh}{HH}{Hard Handover}
\newacronym{hol}{HOL}{Head-of-Line}
\newacronym{hqf}{HQF}{Highest-quality-first}
\newacronym{hss}{HSS}{Home Subscription Server}
\newacronym{http}{HTTP}{HyperText Transfer Protocol}
\newacronym{ia}{IA}{Initial Access}
\newacronym{iab}{IAB}{Integrated Access and Backhaul}
\newacronym{ic}{IC}{Incident Command}
\newacronym{ietf}{IETF}{Internet Engineering Task Force}
\newacronym{ifw}{IFW}{Interference Free Window}
\newacronym{imsi}{IMSI}{International Mobile Subscriber Identity}
\newacronym{imt}{IMT}{International Mobile Telecommunication}
\newacronym{iot}{IoT}{Internet of Things}
\newacronym{ip}{IP}{Internet Protocol}
\newacronym{iq}{IQ}{In-phase and Quadrature}
\newacronym{isi}{ISI}{Inter-Symbol Interference}
\newacronym{itu}{ITU}{International Telecommunication Union}
\newacronym{kpi}{KPI}{Key Performance Indicator}
\newacronym{kvm}{KVM}{Kernel-based Virtual Machine}
\newacronym{lfsr}{LFSR}{Linear Feedback Shift Register}
\newacronym{los}{LOS}{Line-of-Sight}
\newacronym{ls}{LS}{Loosely Synchronised}
\newacronym{lsm}{LSM}{Link-to-System Mapping}
\newacronym{lstm}{LSTM}{Long Short Term Memory}
\newacronym{lte}{LTE}{Long Term Evolution}
\newacronym{lxc}{LXC}{Linux Container}
\newacronym{m2m}{M2M}{Machine to Machine}
\newacronym{mac}{MAC}{Medium Access Control}
\newacronym{mai}{MAI}{Multiple Access Interference}
\newacronym{manet}{MANET}{Mobile Ad Hoc Network}
\newacronym{mano}{MANO}{Management and Orchestration}
\newacronym{mc}{MC}{Multi-Connectivity}
\newacronym{mcc}{MCC}{Mobile Cloud Computing}
\newacronym{mchem}{MCHEM}{Massive Channel Emulator}
\newacronym{mcs}{MCS}{Modulation and Coding Scheme}
\newacronym{mec}{MEC}{Multi-access Edge Computing}
\newacronym{mec2}{MEC}{Mobile Edge Cloud}
\newacronym{mec3}{MEC}{Mobile Edge Computing}
\newacronym{mfc}{MFC}{Mobile Fog Computing}
\newacronym{mi}{MI}{Mutual Information}
\newacronym{mib}{MIB}{Master Information Block}
\newacronym{miesm}{MIESM}{Mutual Information Based Effective SINR}
\newacronym{mimo}{MIMO}{Multiple Input, Multiple Output}
\newacronym{mgen}{MGEN}{Multi-Generator}
\newacronym{ml}{ML}{Machine Learning}
\newacronym{mlr}{MLR}{Maximum-local-rate}
\newacronym[plural=\gls{mme}s,firstplural=Mobility Management Entities (MMEs)]{mme}{MME}{Mobility Management Entity}
\newacronym{mmtc}{mMTC}{Massive Machine-Type Communications}
\newacronym{mmwave}{mmWave}{millimeter wave}
\newacronym{mpdccp}{MP-DCCP}{Multipath Datagram Congestion Control Protocol}
\newacronym{mptcp}{MPTCP}{Multipath TCP}
\newacronym{mr}{MR}{Maximum Rate}
\newacronym{mrdc}{MR-DC}{Multi \gls{rat} \gls{dc}}
\newacronym{mse}{MSE}{Mean Square Error}
\newacronym{mss}{MSS}{Maximum Segment Size}
\newacronym{mt}{MT}{Mobile Termination}
\newacronym{mtd}{MTD}{Machine-Type Device}
\newacronym{mtu}{MTU}{Maximum Transmission Unit}
\newacronym{mumimo}{MU-MIMO}{Multi-user \gls{mimo}}
\newacronym{mvno}{MVNO}{Mobile Virtual Network Operator}
\newacronym{nalu}{NALU}{Network Abstraction Layer Unit}
\newacronym{nas}{NAS}{Network Attached Storage}
\newacronym{nbiot}{NB-IoT}{Narrow Band IoT}
\newacronym{nfv}{NFV}{Network Function Virtualization}
\newacronym{nfvi}{NFVI}{Network Function Virtualization Infrastructure}
\newacronym{nic}{NIC}{Network Interface Card}
\newacronym{nlos}{NLOS}{Non-Line-of-Sight}
\newacronym{now}{NOW}{Non Overlapping Window}
\newacronym{nrdz}{NRDZ}{National Radio Dynamic Zone}
\newacronym{nsf}{NSF}{National Science Foundation}
\newacronym{nsm}{NSM}{Network Service Mesh}
\newacronym[type=hidden]{nr}{NR}{New Radio}
\newacronym{nrf}{NRF}{Network Repository Function}
\newacronym{nsa}{NSA}{Non Stand Alone}
\newacronym{nse}{NSE}{Network Slicing Engine}
\newacronym{nssf}{NSSF}{Network Slice Selection Function}
\newacronym{ntp}{NTP}{Network Time Protocol}
\newacronym{o2i}{O2I}{Outdoor to Indoor}
\newacronym{oai}{OAI}{OpenAirInterface}
\newacronym{oaicn}{OAI-CN}{\gls{oai} \acrlong{cn}}
\newacronym{oairan}{OAI-RAN}{\acrlong{oai} \acrlong{ran}}
\newacronym{oam}{OAM}{Operations, Administration and Maintenance}
\newacronym[plural=\gls{obu}s,firstplural=Onboard Units (OBUs)]{obu}{OBU}{Onboard Unit}
\newacronym{ofdm}{OFDM}{Orthogonal Frequency Division Multiplexing}
\newacronym{olia}{OLIA}{Opportunistic Linked Increase Algorithm}
\newacronym{omec}{OMEC}{Open Mobile Evolved Core}
\newacronym{onap}{ONAP}{Open Network Automation Platform}
\newacronym{onf}{ONF}{Open Networking Foundation}
\newacronym{onos}{ONOS}{Open Networking Operating System}
\newacronym{oom}{OOM}{\gls{onap} Operations Manager}
\newacronym{opnfv}{OPNFV}{Open Platform for \gls{nfv}}
\newacronym[type=hidden]{oran}{O-RAN}{Open \gls{ran}}
\newacronym{orbit}{ORBIT}{Open-Access Research Testbed for Next-Generation Wireless Networks}
\newacronym{os}{OS}{Operating System}
\newacronym{osm}{OSM}{Open Street Map}
\newacronym{oss}{OSS}{Operations Support System}
\newacronym{pa}{PA}{Position-aware}
\newacronym{pase}{PASE}{Prioritization, Arbitration, and Self-adjusting Endpoints}
\newacronym{pawr}{PAWR}{Platforms for Advanced Wireless Research}
\newacronym{pbch}{PBCH}{Physical Broadcast Channel}
\newacronym{pcef}{PCEF}{Policy and Charging Enforcement Function}
\newacronym{pcfich}{PCFICH}{Physical Control Format Indicator Channel}
\newacronym{pcrf}{PCRF}{Policy and Charging Rules Function}
\newacronym{pdcch}{PDCCH}{Physical Downlink Control Channel}
\newacronym{pdcp}{PDCP}{Packet Data Convergence Protocol}
\newacronym{pdsch}{PDSCH}{Physical Downlink Shared Channel}
\newacronym{pdu}{PDU}{Packet Data Unit}
\newacronym{pdp}{PDP}{Power Delay Profile}
\newacronym{pf}{PF}{Proportional Fair}
\newacronym{pgw}{PGW}{Packet Gateway}
\newacronym{phich}{PHICH}{Physical Hybrid ARQ Indicator Channel}
\newacronym{phy}{PHY}{Physical}
\newacronym{pl}{PL}{Path Loss}
\newacronym{pmch}{PMCH}{Physical Multicast Channel}
\newacronym{pmi}{PMI}{Precoding Matrix Indicators}
\newacronym{powder}{POWDER}{Platform for Open Wireless Data-driven Experimental Research}
\newacronym{ppo}{PPO}{Proximal Policy Optimization}
\newacronym{ppp}{PPP}{Poisson Point Process}
\newacronym{prach}{PRACH}{Physical Random Access Channel}
\newacronym{prb}{PRB}{Physical Resource Block}
\newacronym{psnr}{PSNR}{Peak Signal to Noise Ratio}
\newacronym{pss}{PSS}{Primary Synchronization Signal}
\newacronym{pucch}{PUCCH}{Physical Uplink Control Channel}
\newacronym{pusch}{PUSCH}{Physical Uplink Shared Channel}
\newacronym{qam}{QAM}{Quadrature Amplitude Modulation}
\newacronym{qci}{QCI}{\gls{qos} Class Identifier}
\newacronym{qoe}{QoE}{Quality of Experience}
\newacronym{qos}{QoS}{Quality of Service}
\newacronym{qtgui}{QT-GUI}{QT Graphical User Interface}
\newacronym{quic}{QUIC}{Quick UDP Internet Connections}
\newacronym{rach}{RACH}{Random Access Channel}
\newacronym{ran}{RAN}{Radio Access Network}
\newacronym[firstplural=Radio Access Technologies (RATs)]{rat}{RAT}{Radio Access Technology}
\newacronym{rcn}{RCN}{Research Coordination Network}
\newacronym{rec}{REC}{Radio Edge Cloud}
\newacronym{red}{RED}{Random Early Detection}
\newacronym{renew}{RENEW}{Reconfigurable Eco-system for Next-generation End-to-end Wireless}
\newacronym{rf}{RF}{Radio Frequency}
\newacronym{rfc}{RFC}{Request for Comments}
\newacronym{rfr}{RFR}{Random Forest Regressor}
\newacronym{ric}{RIC}{\gls{ran} Intelligent Controller}
\newacronym{rlc}{RLC}{Radio Link Control}
\newacronym{rlf}{RLF}{Radio Link Failure}
\newacronym{rlnc}{RLNC}{Random Linear Network Coding}
\newacronym{rmse}{RMSE}{Root Mean Squared Error}
\newacronym{rnis}{RNIS}{Radio Network Information Service}
\newacronym{rr}{RR}{Round Robin}
\newacronym{rrc}{RRC}{Radio Resource Control}
\newacronym{rrm}{RRM}{Radio Resource Management}
\newacronym{rru}{RRU}{Remote Radio Unit}
\newacronym{rs}{RS}{Remote Server}
\newacronym{rsrp}{RSRP}{Reference Signal Received Power}
\newacronym{rsrq}{RSRQ}{Reference Signal Received Quality}
\newacronym{rss}{RSS}{Received Signal Strength}
\newacronym{rssi}{RSSI}{Received Signal Strength Indicator}
\newacronym{rsu}{RSU}{Road-Side Unit}
\newacronym{rtt}{RTT}{Round Trip Time}
\newacronym{ru}{RU}{Radio Unit}
\newacronym{rw}{RW}{Receive Window}
\newacronym{rx}{RX}{Receiver}
\newacronym{s1ap}{S1AP}{S1 Application Protocol}
\newacronym{sa}{SA}{standalone}
\newacronym{sack}{SACK}{Selective Acknowledgment}
\newacronym{sap}{SAP}{Service Access Point}
\newacronym{sc2}{SC2}{Spectrum Collaboration Challenge}
\newacronym{scef}{SCEF}{Service Capability Exposure Function}
\newacronym{sch}{SCH}{Secondary Cell Handover}
\newacronym{scoot}{SCOOT}{Split Cycle Offset Optimization Technique}
\newacronym{sctp}{SCTP}{Stream Control Transmission Protocol}
\newacronym{sdap}{SDAP}{Service Data Adaptation Protocol}
\newacronym{sd}{SD}{Standard Deviation}
\newacronym{sdk}{SDK}{Software Development Kit}
\newacronym{sdm}{SDM}{Space Division Multiplexing}
\newacronym{sdma}{SDMA}{Spatial Division Multiple Access}
\newacronym{sdn}{SDN}{Software-defined Networking}
\newacronym{sdr}{SDR}{Software-defined Radio}
\newacronym{seba}{SEBA}{SDN-Enabled Broadband Access}
\newacronym{sgsn}{SGSN}{Serving GPRS Support Node}
\newacronym{sgw}{SGW}{Service Gateway}
\newacronym{si}{SI}{Study Item}
\newacronym{sib}{SIB}{Secondary Information Block}
\newacronym{sinr}{SINR}{Signal to Interference plus Noise Ratio}
\newacronym{sip}{SIP}{Session Initiation Protocol}
\newacronym{siso}{SISO}{Single Input, Single Output}
\newacronym{sla}{SLA}{Service Level Agreement}
\newacronym{sm}{SM}{Saturation Mode}
\newacronym{smf}{SMF}{Session Management Function}
\newacronym{smo}{SMO}{Service Management and Orchestration}
\newacronym{sms}{SMS}{Short Message Service}
\newacronym{smsgmsc}{SMS-GMSC}{\gls{sms}-Gateway}
\newacronym{snr}{SNR}{Signal-to-Noise-Ratio}
\newacronym{son}{SON}{Self-Organizing Network}
\newacronym{sptcp}{SPTCP}{Single Path TCP}
\newacronym{srb}{SRB}{Service Radio Bearer}
\newacronym{srn}{SRN}{Standard Radio Node}
\newacronym{srs}{SRS}{Sounding Reference Signal}
\newacronym{ss}{SS}{Synchronization Signal}
\newacronym{sss}{SSS}{Secondary Synchronization Signal}
\newacronym{st}{ST}{Spanning Tree}
\newacronym{svc}{SVC}{Scalable Video Coding}
\newacronym{tb}{TB}{Transport Block}
\newacronym{tcp}{TCP}{Transmission Control Protocol}
\newacronym{tdd}{TDD}{Time Division Duplexing}
\newacronym{tdm}{TDM}{Time Division Multiplexing}
\newacronym{tdma}{TDMA}{Time Division Multiple Access}
\newacronym{tfl}{TfL}{Transport for London}
\newacronym{tfrc}{TFRC}{TCP-Friendly Rate Control}
\newacronym{tft}{TFT}{Traffic Flow Template}
\newacronym{tgen}{TGEN}{Traffic Generator}
\newacronym{tip}{TIP}{Telecom Infra Project}
\newacronym{tm}{TM}{Transparent Mode}
\newacronym{to}{TO}{Telco Operator}
\newacronym{toa}{ToA}{Time of Arrival}
\newacronym{tr}{TR}{Technical Report}
\newacronym{trp}{TRP}{Transmitter Receiver Pair}
\newacronym{ts}{TS}{Technical Specification}
\newacronym{tti}{TTI}{Transmission Time Interval}
\newacronym{ttt}{TTT}{Time-to-Trigger}
\newacronym{tx}{TX}{Transmitter}
\newacronym{uas}{UAS}{Unmanned Aerial System}
\newacronym{uav}{UAV}{Unmanned Aerial Vehicle}
\newacronym{udm}{UDM}{Unified Data Management}
\newacronym{udp}{UDP}{User Datagram Protocol}
\newacronym{udr}{UDR}{Unified Data Repository}
\newacronym{ue}{UE}{User Equipment}
\newacronym{uhd}{UHD}{\gls{usrp} Hardware Driver}
\newacronym{ul}{UL}{Uplink}
\newacronym{um}{UM}{Unacknowledged Mode}
\newacronym{uml}{UML}{Unified Modeling Language}
\newacronym{upa}{UPA}{Uniform Planar Array}
\newacronym{upf}{UPF}{User Plane Function}
\newacronym{urllc}{URLLC}{Ultra Reliable and Low Latency Communication}
\newacronym{usa}{U.S.}{United States}
\newacronym{usim}{USIM}{Universal Subscriber Identity Module}
\newacronym{usrp}{USRP}{Universal Software Radio Peripheral}
\newacronym{utc}{UTC}{Urban Traffic Control}
\newacronym{vim}{VIM}{Virtualization Infrastructure Manager}
\newacronym{vm}{VM}{Virtual Machine}
\newacronym{vnf}{VNF}{Virtual Network Function}
\newacronym{volte}{VoLTE}{Voice over \gls{lte}}
\newacronym{voltha}{VOLTHA}{Virtual OLT HArdware Abstraction}
\newacronym{vr}{VR}{Virtual Reality}
\newacronym{vran}{vRAN}{Virtualized \gls{ran}}
\newacronym{vss}{VSS}{Video Streaming Server}
\newacronym{wbf}{WBF}{Wired Bias Function}
\newacronym{wf}{WF}{Wired-first}
\newacronym{wi}{WI}{Wireless InSite}
\newacronym{wlan}{WLAN}{Wireless Local Area Network}
\newacronym{pnf}{PNF}{Physical Network Function}
\newacronym{drl}{DRL}{Deep Reinforcement Learning}
\newacronym{mtc}{MTC}{Machine-type Communications}
\newacronym{v2x}{V2X}{Vehicle-to-everything}
\newacronym{cast}{CaST}{Channel emulation scenario generator and Sounder Toolchain}
\newacronym{gui}{GUI}{Graphical User Interface}
\newacronym{ups}{UPS}{Uninterruptible Power Supply}
\newacronym{ota}{OTA}{Over-the-Air}
\newacronym{hitl}{HITL}{hardware-in-the-loop}
\newacronym{soc}{SoC}{System-on-Chip}
\newacronym{eeg}{EEG}{electroencephalogram}
\newacronym{ieeg}{iEEG}{intracranial electroencephalogram}
\newacronym{ecg}{ECG}{electrocardiogram}
\newacronym{fph}{FPH}{false positive per hour}
\newacronym{cnn}{CNN}{Convolutional Neural Network}
\newacronym{ban}{BAN}{Body Area Network}
\newacronym{roc}{ROC}{Receiver Operating Characteristic Curve}
\newacronym{auc}{AUC}{Area Under the Curve}
\newacronym{imd}{IMD}{Implantable Medical Devices}
\newacronym{rns}{RNS}{Responsive Neurostimulation}
\newacronym{vns}{VNS}{Vagus Nerve Stimulation}

\usepackage{tikz}
\usepackage{pgfplots}
\usepackage{caption}

\pgfplotsset{compat=newest}
\pgfplotsset{plot coordinates/math parser=false}
\newlength\fheight
\newlength\fwidth
\usetikzlibrary{plotmarks,patterns,decorations.pathreplacing,backgrounds,calc,arrows,arrows.meta,spy,matrix,scopes}
\usepgfplotslibrary{patchplots,groupplots}
\usepackage{tikzscale}

\usepackage{algorithm}
\usepackage{algorithmicx}
\usepackage{algcompatible}
\usepackage[noend]{algpseudocode}

\usepackage{tikzpagenodes,etoolbox}
\usetikzlibrary{calc}
\usepackage[contents={}]{background}
\AddEverypageHook{%
\ifnumequal{\thepage}{1}{%
    \tikz[remember picture,overlay]{%
        \node[draw,
        minimum width=1.03\textwidth,
        text width=1.02\textwidth,
        font=\footnotesize
        ]
        at ($(current page header area)$)
        {%
        This paper has been accepted for publication in the Proceedings of the IEEE 20th International Conference on Body Sensor Networks (BSN). This is the author's accepted version of the article. The final version published by IEEE is A. Saeizadeh, D. Schonholtz, J. S. Neimat, P. Johari, T. Melodia, “A Multi-Modal Non-Invasive Deep Learning Framework for Progressive Prediction of Seizures,” in Proceedings of the IEEE 20th International Conference on Body Sensor Networks (BSN), October 2024.
        };
    }%
}{}
}

\begin{document}

\title{A Multi-Modal Non-Invasive Deep Learning Framework for Progressive Prediction of 
Seizures}
\author{\IEEEauthorblockN{Ali Saeizadeh$^\dagger$, Douglas Schonholtz$^\dagger$, Joseph S. Neimat$^*$, Pedram Johari$^\dagger$, Tommaso Melodia$^\dagger$}

\IEEEauthorblockA{$^\dagger$Institute for the Wireless Internet of Things, Northeastern University, Boston, MA, U.S.A.\\
$^*$University of Louisville, Louisville, KY, U.S.A.\\
E-mail: $^\dagger$\{saeizadeh.a, schonholtz.d, p.johari, melodia\}@northeastern.edu, $^*$joseph.neimat@uoflhealth.org\\
}
}

\maketitle


\begin{abstract}
This paper introduces an innovative framework designed for progressive (granular in time to onset) prediction of seizures through the utilization of a \gls{dl} methodology based on non-invasive multi-modal sensor networks. Epilepsy, a debilitating neurological condition, affects an estimated 65 million individuals globally, with a substantial proportion facing drug-resistant epilepsy despite pharmacological interventions. To address this challenge, we advocate for predictive systems that provide timely alerts to individuals at risk, enabling them to take precautionary actions. Our framework employs advanced \gls{dl} techniques and uses personalized data from a network of non-invasive \gls{eeg} and \gls{ecg} sensors, thereby enhancing prediction accuracy. The algorithms are optimized for real-time processing on edge devices, mitigating privacy concerns and minimizing data transmission overhead inherent in cloud-based solutions, ultimately preserving battery energy. Additionally, our system predicts the countdown time to seizures (with 15-minute intervals up to an hour prior to the onset), offering critical lead time for preventive actions. Our multi-modal model achieves 95\% sensitivity, 98\% specificity, and 97\% accuracy, averaged among 29 patients.
\end{abstract}

\maketitle

\section{Introduction}\label{sec:intro}


Epilepsy is one of the most common and deadly neurological diseases with around 50 million people diagnosed worldwide and a risk of death three times higher than the general population \cite{who}. In the U.S. alone, 150,000 people are diagnosed with epilepsy every year and about 2.3 million adults and more than 450,000 children and adolescents currently live with this life-crippling condition. \$15.5B  is the U.S. estimate in annual medical expenditures and lost or reduced earning and productivity due to epilepsy \cite{NIH_epilepsy_and_seizures}. Although most patients diagnosed with epilepsy respond well to pharmaceutical drugs to treat epilepsy, there is still a large 30\% to 40\% of patients that suffer from drug resistant epilepsy and cannot be treated with traditional drugs \cite{xue2019risk}. Therefore, there is a clear need for epilepsy treatment that goes beyond pharmacological solutions. 

\begin{figure}[ht]
\centering
\includegraphics[width=0.35\textwidth]{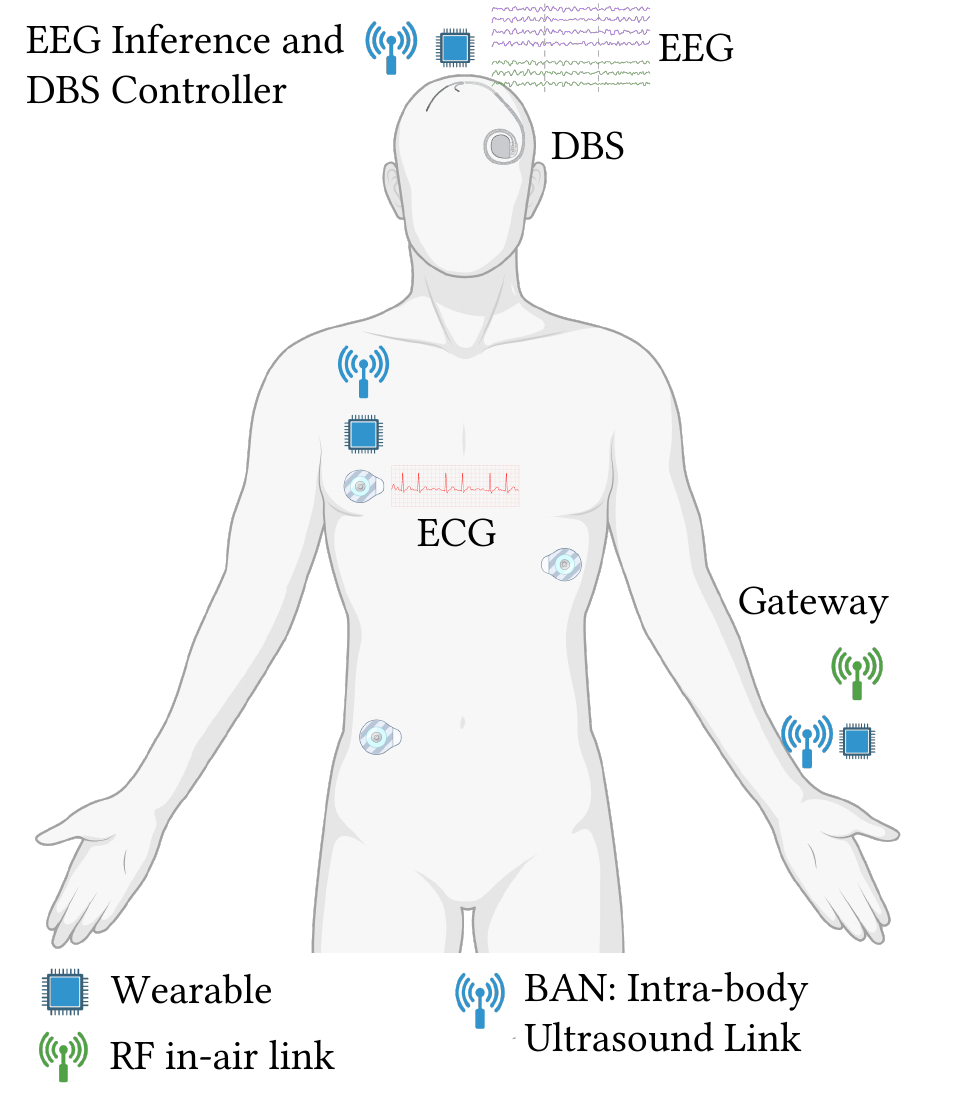}\caption{Proposed ultrasonically connected \gls{ban}. The gateway receives the classification results from the EEG and ECG nodes that execute deep learning algorithms and sends commands to the \gls{dbs} through its external controller using ultrasonic signals (This figure is generated by BioRender).}
\label{fig:system}
\vspace{-15pt}
\end{figure}


Seizure identification and prediction can promise substantial treatment benefits for patients suffering from epilepsy. Seizure identification has, for more than a decade, been incorporated in neuromodulation devices with considerable success \cite{RN22260, RN22261, RN22262}. The benefits are twofold: (i) the systems have been designed to initiate stimulation in response to signal changes in \gls{eeg} (as in the \gls{rns} device) \cite{RN22260, RN22262} or in \gls{ecg} (as with recent \gls{vns} devices) \cite{RN22261}. In both cases, this has proved an effective strategy to achieve seizure reduction; (ii) the institution of seizure detection and long-term tracking, as in the \gls{rns} device, has provided an objective measure of treatment success, or lack thereof. This is true not just for the stimulation provided by the device itself, but also for changes in medication or the implementation of other medical and surgical treatments~\cite{RN22267}. 

Seizure prediction, however, opens novel treatment opportunities. By anticipating probability of seizure, future devices may alter their stimulation pattern or amplitude to reduce this probability, perhaps ultimately preventing seizures altogether. Novel devices that inform the patient of this increased probability may enable patients to alert care-givers or avoid activities that could be dangerous during seizure.

Two persisting challenges in seizure prediction are (i) training datasets are largely imbalanced due to the scarce occurrence of seizure vs non-seizure periods; and (ii) classifying the per-seizure and non-seizure indicators with high accuracy long in advance of the seizure onset (up to 1-hour) is not trivial.
The first challenge arises due to the infrequency of seizures in patient recordings, making data-driven methods prone to bias. Prior research has addressed this issue through under-sampling the non-seizure periods. 
On the second note, for the pre-seizure samples that are further in advance of the onset (one hour), the classification often leads to larger false negatives (low sensitivity), while for the non-seizure samples the false positive tends to be larger as it gets closer to the pre-seizure period (low specificity). 

Although a multitude of studies have utilized individual biological time-series signals such as \gls{eeg}, \gls{ieeg}, or \gls{ecg} to develop \gls{dl}-based solutions for seizure prediction \cite{park2011seizure, billeci2018patient, uvaydov2022aieeg}, the potential advantages of integrating multiple modalities remain largely unexplored. The implementation of a multi-modal approach could potentially enhance the robustness of the prediction models, thereby reducing variance and improving overall performance \cite{schulze2022seizure}.

In this study we leverage the synergy of non-invasive \gls{eeg} and \gls{ecg} signals to enhance the accuracy of seizure prediction. This combined approach not only holds promise for advancing the development of next-generation closed-loop neuromodulation systems,
but also aligns with the growing ubiquity of wearable devices. The latter could further enable the integration of non-invasive cardiac-based signals into seizure prediction systems, thereby broadening the scope and effectiveness of these technologies \cite{RN22251, RN22253}. Furthermore, for the first time, we implement a multi-class prediction algorithm that can predict the seizure in a progressive fashion, i.e., in 15~min time intervals prior to the seizure onset, and up to one hour in advance. 



We address the aforementioned challenges by introducing a personalized framework for multi-modal seizure prediction, using non-invasive \gls{eeg} and \gls{ecg} signals. We design a \gls{dl} algorithm to process both types of sensor readings, and we study efficient methods for integrating the multi-modal classifications derived from each sensor to achieve the highest level of accuracy. This integration aims at classifying progressive pre-seizure (\textit{preictal}) periods --divided in 15~min time intervals within an hour--, and the non-seizure (\textit{interictal}) period.

\section{System Overview}  
\label{sec:sensor-network}

As shown in Fig.~\ref{fig:system}, the proposed solution consists a \gls{ban} of both wearable and implantable nodes performing multiple functions, such as sensing, computation, communication, and actuation. 
The system operates in a closed-loop configuration, utilizing sensors to monitor physiological signals, processing these signals with \gls{dl} capabilities, and providing feedback to an implanted \gls{dbs} or a preemptive alert system. The gateway device acts as an interface for the patient to interact with the system, offering continuous brain activity monitoring, seizure alerts, and the ability to adjust \gls{dbs} settings. This setup ensures that both the patient and medical professionals can access and manage the therapy remotely, enhancing the system's efficiency and responsiveness.

\gls{ban} consists of four key components: the \gls{eeg} classifier, \gls{ecg} classifiers, gateway, and \gls{dbs} implant. The \gls{eeg} and \gls{ecg} nodes locally process sensor data using \gls{dl} algorithms and transmit the intermediate results to the gateway. The gateway integrates these results, as described in Sec.\ref{sec:dl-system}, to make a final decision on seizure prediction and subsequently sends stimulation commands to the \gls{dbs} implant. The entire inference process runtime on the Xilinx KV260 platform averages 20 ms, ensuring rapid decision-making. This wireless network utilizes ultrasonic communication to maintain reliable data transmission and coordination between nodes. Each node is specifically designed with tailored computational, memory, communication, and energy specifications to support its dedicated functions, optimizing overall system performance and enabling timely medical interventions. The feasibility of the system has been previously discussed in our earlier work\cite{saeizadeh2023wons}.

\section{\gls{dl}-Based Seizure Prediction} 
\label{sec:dl-system}

\begin{figure}
    \centering
    \includegraphics[width=0.45\textwidth]{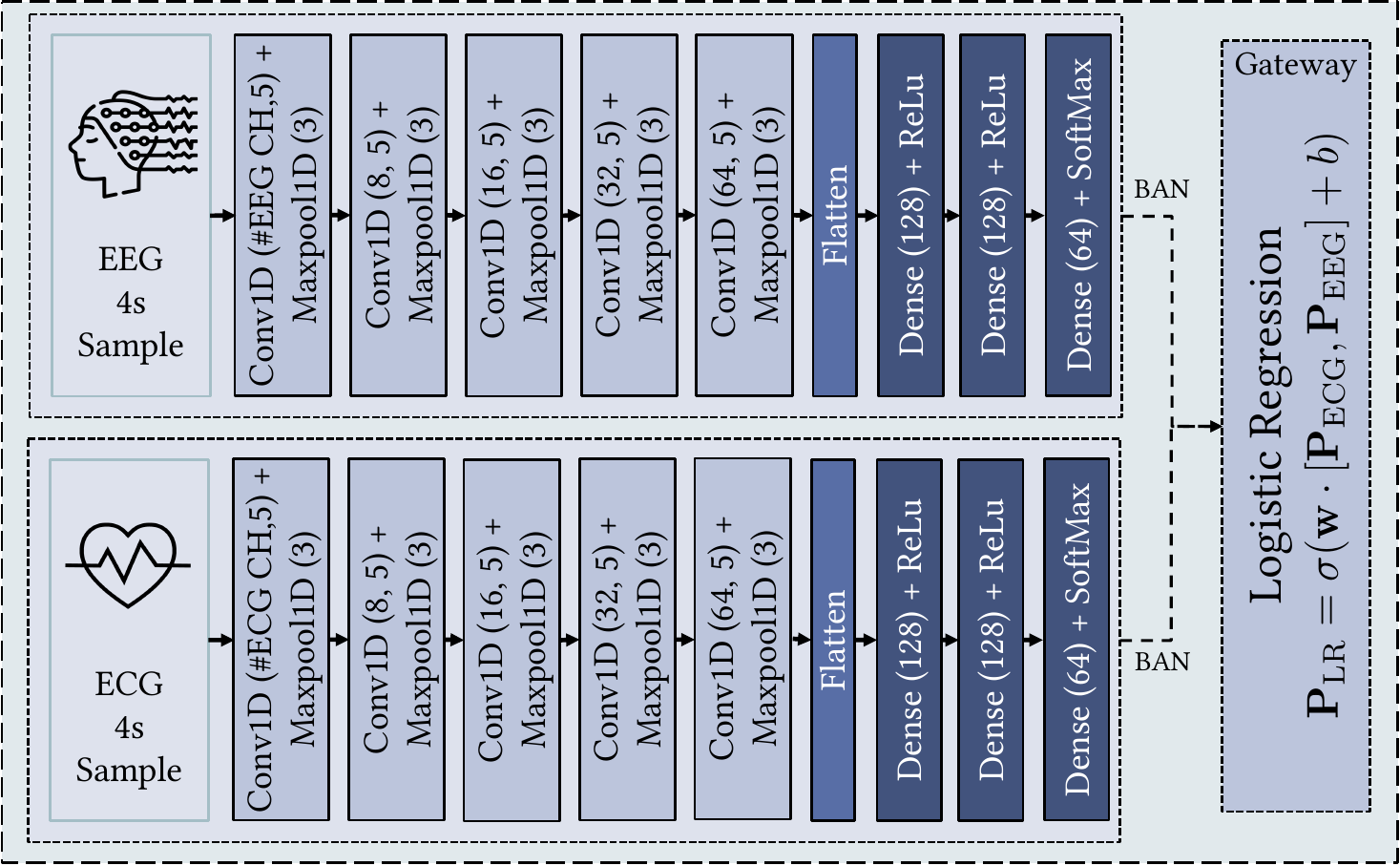}
    \caption{Deep Learning Model and Prediction System Structure}
    \label{fig:dl_model}
    \vspace{-15pt}
\end{figure}

In this paper, we leverage a comprehensive dataset produced as part of the EPILEPSIAE project \cite{ihle2012epilepsiae}. This dataset contains a wide variety of biosignals collected from 275 patients diagnosed with focal epilepsy. The data, gathered between 2009 and 2012 from three esteemed European centers, is characterized by continuous long-term recordings, averaging 165 hours per patient and an average of 9.8 seizures per patient. Among the 275 patients, 29 have non-invasive data comprising both single-channel \gls{ecg} recordings from the chest and surface \gls{eeg} data in the 10–20 system. In our study, we utilize all the available data from these 29 patients to ensure a comprehensive analysis of seizure prediction using non-invasive methods. While the influence of noise in \gls{eeg} and \gls{ecg} signals generated by daily activities on detection results is acknowledged, this aspect will be investigated in future work.

As our focus in this paper in on predicting seizure, we ignore seizure (\textit{ictal}) periods during which a seizure is occurring. Then, we define five labels for the periods of the recordings, 4 of which are the labels for the pre-seizure (\textit{preictal}) intervals with specific periods of time prior to a seizure, which are set to 15, 30, 45, or 60 minutes before seizure onset. The last label, indicates the non-seizure (\textit{interictal}) class, associated to the periods which are neither in the \textit{ictal} period nor close to it (the threshold is set to be over an hour in advance of the onset). Utilizing an end-to-end model enables us to avoid the overhead associated with data pre-processing, thereby optimizing power consumption and minimizing the physical size of the system. Our objective for the prediction model is to maintain lightness of the model, so it can be embedded on a wearable/implantable medical device.

\begin{figure*}[!b]
    \vspace{-10pt}
     \centering
     \begin{subfigure}{0.3\textwidth}
         \centering
         \includegraphics[width=\textwidth]{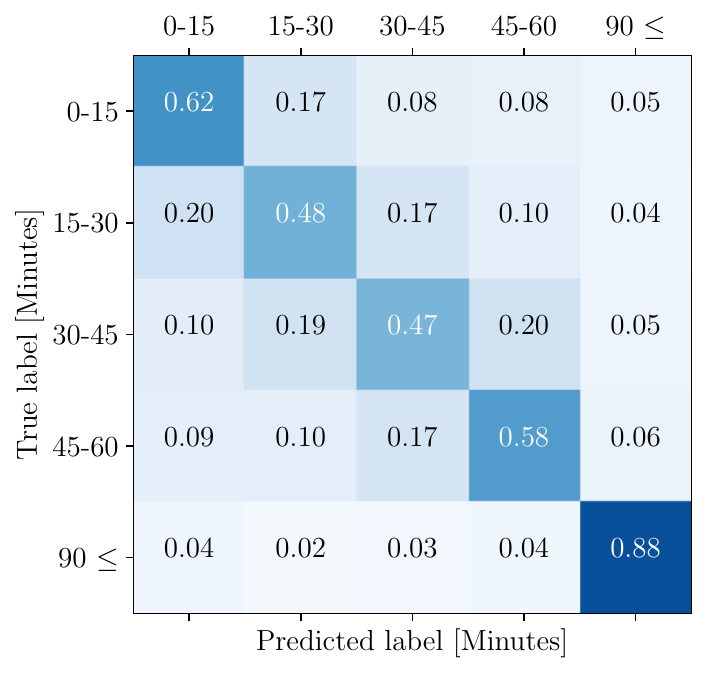}
         \caption{ECG}
         \label{fig:cm_ecg}
     \end{subfigure}
     \hfill
     \begin{subfigure}{0.3\textwidth}
         \centering
         \includegraphics[width=\textwidth]{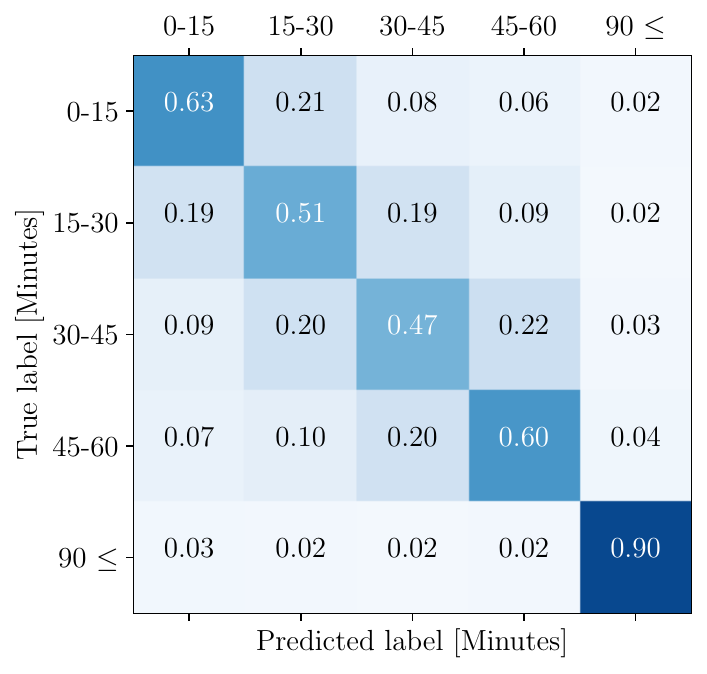}
         \caption{EEG}
         \label{fig:cm_eeg}
     \end{subfigure}
     \hfill
     \begin{subfigure}{0.3\textwidth}
         \centering
         \includegraphics[width=\textwidth]{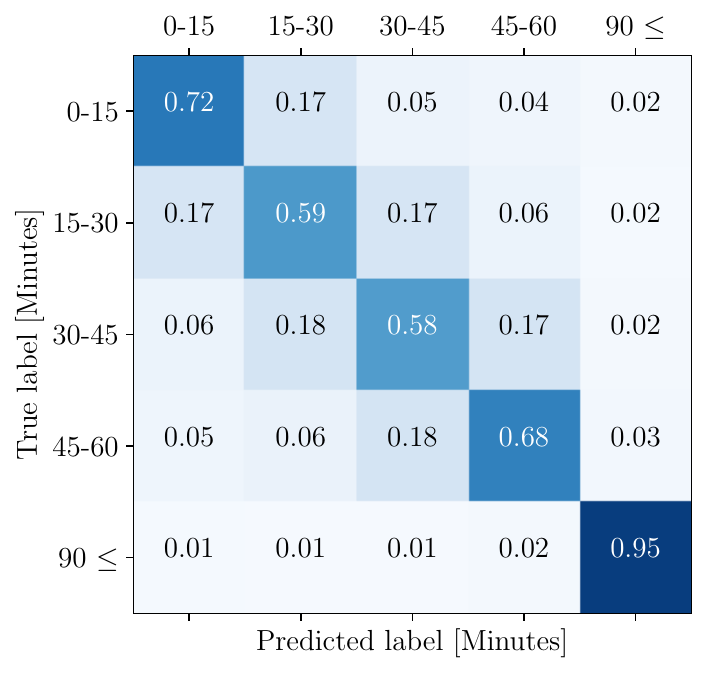}
         \caption{Combined ECG and EEG}
         \label{fig:cm_comb}
     \end{subfigure}
     \caption{Confusion Matrices on the test dataset for different modalities. (averaged among the patients)}
     \label{fig:cm}
\end{figure*}

\subsection{Deep Learning Model}
Our \gls{ecg}/\gls{eeg} analysis methodology, illustrated in Fig.~\ref{fig:dl_model}, starts with batch normalization of raw samples. These samples are then processed through four 1-D CNN and Max-pooling blocks for feature extraction, where the number of channels in each block matches the number of input channels. The extracted features are flattened and passed through three dense layers for binary classification, utilizing ReLU and SoftMax activation functions to get the probabilities ($\mathbf{P}_{\text{EEG}}$ and $\mathbf{P}_{\text{ECG}}$) for each class as shown in (\ref{eqn:cnn}). Despite achieving high accuracy, the model is designed to minimize the number of weights and biases, thereby reducing computational cost and making it suitable for wearable devices.
\begin{equation}
\begin{aligned}
& \mathbf{P}_{\text{EEG}}(Y = k \mid X_{\text{EEG}}, \mathbf{w}_{\text{EEG}}), \;  \\
& \mathbf{P}_{\text{ECG}}(Y = k \mid X_{\text{ECG}}, \mathbf{w}_{\text{ECG}}), \\
& k \in \{0-15, 15-30, 30-45, 45-60, >90\}\; \text{minutes} \\
\end{aligned}
\label{eqn:cnn}
\end{equation}

We individually train the ECG and EEG models for each patient using the Focal Loss function, as introduced by \cite{lin2017focal} and utilized in our prior work~\cite{saeizadeh2023wons} to find the optimal weights and biases, $\mathbf{w}_{\text{ECG}}$ and $\mathbf{w}_{\text{EEG}}$. This loss function effectively addresses data imbalance, thereby enhancing the performance of the models. Additionally, a logistic regression model is trained concurrently with the deep learning models.

\subsection{Combiner Model}
To integrate the results, we extract the Softmax layers' outputs from the EEG and ECG models, containing the probabilities associated with the predicted labels. These probabilities are then fed into a dedicated logistic regression model, referred to as the combiner model, to yield the final results. The combiner model takes 10 inputs—five probabilities from each model ($\mathbf{P}_{\text{ECG}}$ and $\mathbf{P}_{\text{EEG}}$) in (\ref{eqn:comb}). The probabilities are provided with four-digit precision to remain within the bandwidth constraints of the \gls{ban} system during both training and operation.

\begin{equation}
\begin{aligned}
   \mathbf{X} = [\mathbf{P}_{\text{EEG}}(Y = k \mid X_{\text{EEG}}, \mathbf{w}_{\text{EEG}}), & \\ 
   \mathbf{P}_{\text{ECG}}(Y = k \mid X_{\text{ECG}}, \mathbf{w}_{\text{ECG}})] 
\label{eqn:comb}
\end{aligned}
\end{equation}

$\mathbf{X}$ undergoes training to optimize its weights, ensuring the best possible combination of results
Logistic Regression output. The logistic regression probability for class \( k \) is:
\begin{equation}
    \begin{aligned}
        P_{\text{LR}}(Y = k \mid \mathbf{X}) = 
        \frac{e^{\mathbf{w}_k \cdot \mathbf{X} + b_k}}{\sum_{j=1}^K e^{\mathbf{w}_j \cdot \mathbf{X} + b_j}},
    \end{aligned}
\end{equation}
\noindent where \( \mathbf{w}_k \) and \( b_k \) are the weight vector and bias for class \( k \). The final predicted class \( \hat{Y} \) is:
\begin{equation}
   \hat{Y} = \arg\max_{k} P_{\text{LR}}(Y = k \mid \mathbf{X}) 
\end{equation}

\section{Experimental Results} \label{sec:exp_results}
All reported results are evaluated using 5-fold cross-validation. In each fold, the dataset is split into 80\% for training and 20\% for testing, with 10\% of the training set further allocated for validation. The performance metrics represent the average efficacy of the model across all folds, specifically on the testing datasets. This comprehensive approach ensures a robust assessment of the model's performance across varying data subsets.

The averaged  normalized confusion matrices among all the patients, shown in Fig.~\ref{fig:cm}, include the \gls{ecg}, \gls{eeg}, and combined models. As observed, the system demonstrates strong performance on non-seizure samples (associated to the ``$90$$\leq$'' class). Although the model exhibits lower accuracy for each pre-seizure class (divided in 15~min time intervals), it is evident that most errors occur with adjacent labels. This suggests that adjacent classes can contribute in boosting the accuracy of one another.

\begin{figure}

    \centering
    \includegraphics[width=0.45\textwidth]{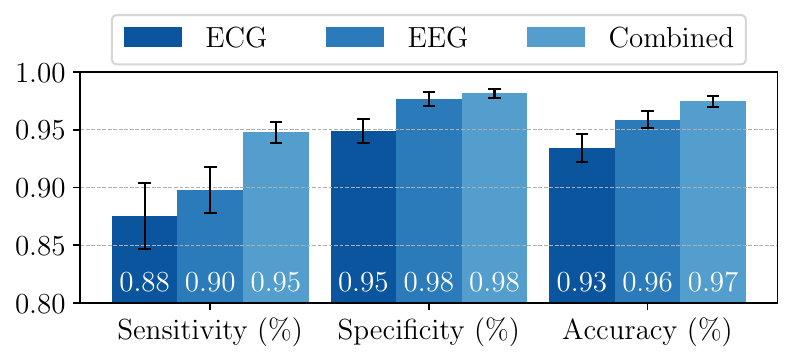}
    \caption{Average metrics among all the patients for different models.}
    \label{fig:results}
    \vspace{-10pt}
\end{figure}

\begin{figure}
    \centering
         \includegraphics[width=0.45\textwidth]{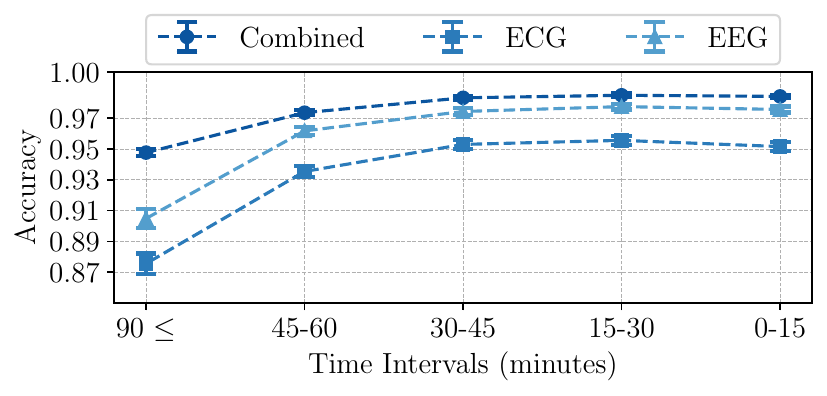}
         \caption{Accuracy trend over time intervals leading up to seizure onset.}
        \label{fig:trend}
        \vspace{-10pt}
\end{figure}

To assess the effectiveness of our model in binary classification, particularly in identifying whether a patient is within 60 minutes of seizure onset, we classify all labels under 60 minutes as pre-seizure and all others as non-seizure. We utilize a range of metrics for a comprehensive analysis, including sensitivity, specificity, and accuracy, as depicted in Fig.~\ref{fig:results}. The results presented are averaged across all patients. Additionally, we plotted the accuracy trend for binary classification over time in Fig.~\ref{fig:trend}, demonstrating the expected improvement in accuracy as the predicted time of seizure onset approaches.

\section{Conclusion}
In conclusion, this paper introduces a robust system for seizure prediction utilizing sensor networks, \gls{dl} models, and advanced data curation techniques. The architecture incorporates \gls{eeg} and \gls{ecg} classifiers, connected to a gateway for real-time decision-making. Leveraging a large-scale dataset from the EPILEPSIAE project, the system addresses class imbalance challenges through a focal loss function in the \gls{dl} model. The proposed sensor network, provides a framework to interconnect wearable, implantable, and brain-stimulator nodes, forming a closed-loop system for effective monitoring and intervention. Additionally, our system is capable to predict the time to seizures in multiple classes (with 15 minutes intervals up to an hour prior to the onset), offering critical lead time for preventive actions. Experimental results showcase the system's outstanding sensitivity, specificity, and accuracy ($>$97\%) compared to the state-of-the-art.

\bibliographystyle{ieeetr}
\bibliography{simple_base}

\end{document}